# A Dual-Port 8-T CAM-Based Network Intrusion Detection Engine for IoT

Dai Li, *Student Member, IEEE*, Kaiyuan Yang, *Member, IEEE*

*Abstract*— This work presents an energy- and memory-efficient pattern-matching engine for a network intrusion detection System (NIDS) in the Internet of Things. Tightly coupled architecture and circuit co-designs are proposed to fully exploit the statistical behaviors of NIDS pattern matching. The proposed engine performs pattern matching in three phases, where the phase-1 prefix matching employs reconfigurable pipelined automata processing to minimize memory footprint without loss of throughput and efficiency. The processing elements utilize 8-T content-addressable memory (CAM) cells for dual-port search by leveraging proposed fixed-1s encoding. A 65-nm prototype demonstrates best-in-class 1.54-fJ energy per search per pattern byte and 0.9-byte memory usage per pattern byte.

## I. INTRODUCTION

A network intrusion detection system (NIDS) is highly desired in the Internet of Things (IoT) and wireless sensor networks, which face security issues due to heterogeneous and diverse devices, networks, and applications. A signature-based NIDS, such as Snort [1] and ClamAV [2], examines every packet, both header and payload, against expert crafted rules. Deploying NIDS capabilities on edge devices provides better traffic coverage, scalable computing resource, and higher flexibility in rule deployments than a central NIDS at a hub. But the major barrier is the limited energy budget for edge devices.

Matching the full payload of packets is more powerful and effective in finding malicious packets than header matching and filtering. This process is often referred to as deep packet inspection (DPI). DPI compares the streaming packet against variable-length patterns that may appear at any position. Many DPI rules have additional options including case sensitivity, wildcard matching, and specified distance between the occurrences of multiple patterns. The enormous number of patterns requires high parallelism and bandwidth to process pattern matching. The irregular memory access patterns of DPI make implementing it in von Neumann architectures inefficient.

Pattern matching is the core task in signature-based NIDS, which finds broad applications in security, database, and machine learning. Previous research has explored various hardware and algorithm optimizations, but all have faced trade-offs between energy and memory efficiency (Fig. 1b). Recently, memory centric architectures [3][4] have been proposed to handle the pattern matching problem using deterministic finite automaton (DFA)- and non-deterministic finite automaton (NFA)- based engines, for example, Micron's Automata Processor [5] is a silicon solution implementing State Transition Elements (STE) in DRAM aiming at line speed processing but comes with high power cost of tens of Watts. SRAM-based pattern matching engines [6][7] still suffer from area overhead due to the redundancy in most STEs. Content-addressable memory (CAM) is widely used in pattern matching for its parallel search capability. Previous studies have implemented brute-force [8], Wu-Manber [9] and Aho-Corasick (AC) pattern matching algorithms [10] in TCAM. They are high in throughput but suffer from large searching energy and a large (T)CAM footprint.

In order to improve the efficiency of edge NIDS in terms of energy, memory, and area, this paper presents a CAM-based pattern matching engine with tightly coupled architecture and circuit innovations that

Dai Li and Kaiyuan Yang are with Rice University, Houston, TX 77005 USA.

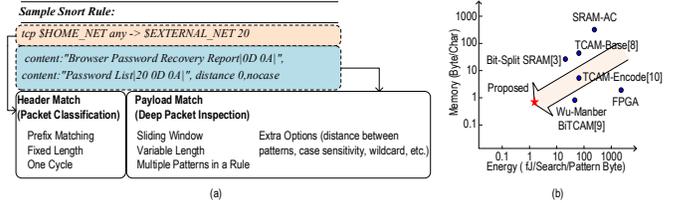

Fig. 1. (a) A Snort rule example, (b) Performance comparison of proposed work with prior arts.

exploits the statistical behaviors of pattern matching: 1) three-phase architecture; 2) memory efficient reconfigurable pipelined AC structure; 3) fine-grained CAM row enabling, power gating and clock gating; 4) dual-port 8-T CAM cell for Phase-1 processing elements (PE) and 5) 7-T CAM cell capable of single-port wildcard matching for Phase-2, enabled by a fixed-1s encoding scheme.

## II. RECONFIGURABLE THREE-PHASE NIDS ARCHITECTURE

Matching NIDS rules in the proposed system is performed in three phases to leverage the statistical characteristics of pattern hit rate for power savings (Fig. 2). Phase-1 matches the prefix using an AC algorithm with a reconfigurable prefix length (depth). The AC algorithm is chosen because of its constant throughput and hardware optimization opportunities. Phase-2 matches the rest of the characters only after Phase-1 has matched the prefix. Phase-3 records the patterns matched in Phase-2 and checks for possible matches of multi-pattern and long pattern rules whenever a new match arrives. A Rule Table and a Partial Hit Table are used in Phase-3 to store transition rules for multi-pattern and long patterns and to record existing hits. It will report a "matched rule" to the user when the last pattern or "suffix" of a rule is matched. Since Phase-3 has probability lower than $2^{-150}$ of usage in most cases, it is implemented in software and off-chip for trade-offs among speed, energy and area. The 3-Phase architecture and the algorithm implemented on the microprocessor are presented in Fig. 3.

The AC algorithm is based on a Trie-style finite automata composed of states and transitions, and is constructed from the pattern library (Fig. 4a). A Trie or prefix tree, is a data structure that stores the prefixes of patterns as a dictionary in a tree for the streaming-in data to search for a match. The finite automata reads one input character per cycle and transits from the current state to the next state based on the current input. There are two types of transitions, forward and backward ones. Forward transitions handle regular prefix matching. The depth of a state is defined as its distance to the root state in forward transitions. Backward transitions occur only when no valid forward transitions exist. They are used to find other states with a shorter prefix matching the suffix of the currently matched string. Conventional CAM-based AC designs [10] store {current state, input character} in CAM and

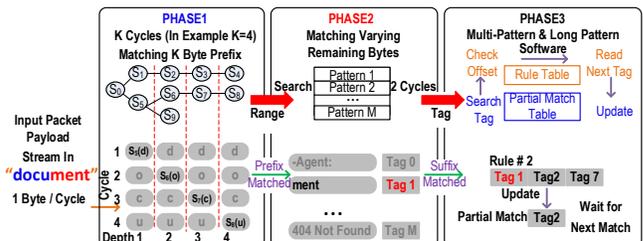

Fig. 2. Proposed 3-Phase pattern matching workflow with an example.

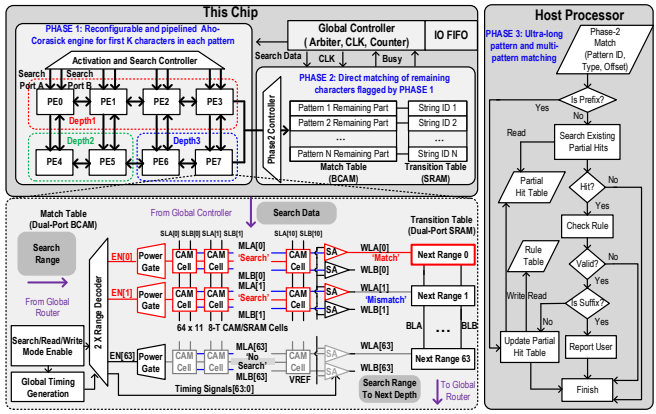

Fig. 3. Block diagram of proposed network intrusion detection system (NIDS) and processing element.

{next state} in the corresponding SRAM entries. Packets streamed in are searched in the CAM for a valid transition and the index for the next state. In this approach memory usage is directly proportional to the number of transitions. Fig. 4b summarizes the transition statistics on a ClamAV ruleset [10] and an IoT-related subset of Snort. Typically, there are many more backward transitions than forward ones because the starting point of a pattern in the input is unknown. Two main drawbacks of conventional CAM-based AC design are: 1) high memory storage due to the number of transitions exponentially increasing with the number of rules, and 2) high power consumption due to a parallel CAM search and high memory volume caused by 1).

While various encoding schemes were proposed to reduce backward transitions, all coding schemes required TCAMs and increase the width of state indices [10]. To completely eliminate backward transitions, we propose dividing the large match/transition table into PE clusters (Fig. 3) that store transitions at different depths in AC Trie and pipeline the execution of PE clusters. Each cluster maintains a current searching state and takes the same input byte in each cycle, as shown in Fig. 5a. As a result of pipelining, no backward transition is needed because there is always a shallower depth already at the destination state of the backward transition. While pipelining increases the total number of CAM searches by a few times, it reduces the number of transitions and the size of tables by orders of magnitude according to Fig. 4b. Furthermore, the clustering of PEs in Phase-1 is designed to be reconfigurable in terms of the number of pipeline stages and the number of PEs in each stage, which enables optimal efficiency under different rulesets and traffic conditions. The first stage in Phase-1 is implemented with an two-port SRAM, instead of the CAM-based PE, because no state matching is needed.

The searching energy is further improved by fine-grained row enabling in the CAMs, based on the observation that only transitions to "children states" of the current state are valid candidates for which to search. Activating only the relevant rows in CAMs (Fig. 6) not only saves energy by up to 35% by spending less energy on match lines, but also reduces the width of CAM entries because the index of the current state is already contained in the range information. Thus, the CAM only stores {input character}, and the SRAM stores {next range} instead of {next state}. {next range} consists of a global PE ID to identify the PE for the next depth's search and a local UP/DN range pointing to the candidate rows for searching next transitions. Based on the global PE ID, a global router activates corresponding PEs and Phase-2 banks, and passes the detailed UP/DN range to them for the next search. Moreover, the row enabling mechanism prevents unnecessary CAM searches when there is no valid state at any depth.

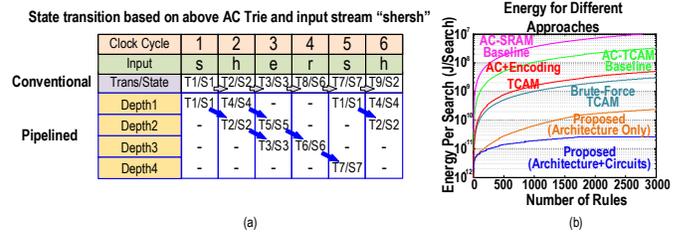

Fig. 5. (a) State and transition of conventional and pipelined AC algorithm on an example, (b) Simulated energy efficiency versus number of rules using different approach.

Phase-2 is implemented with a single wide CAM-SRAM table without pipelining, because of the low activation rate of Phase-2. The CAM in Phase-2 supports wildcard matching for patterns with "don't care" bytes and of different lengths. CAM row enabling is still desired for Phase-2 CAM to save match line power. If a pattern is matched in Phase-2, the entire system is clock-gated to skip comparing the suffix of the matched pattern to save power (measured results in Fig. 12a).

We simulated the scaling of search energy over the ruleset size. Our proposed design with only architectural optimization (not including row enabling) and the complete co-designed system are compared with conventional methods in Fig. 5b, using the same CAM/SRAM energy models extracted from post-layout simulations. The energy of the proposed designs are by orders of magnitude lower than that of the baselines, and show much better scalability over the rule size. This is mainly because each search is confined within a small range of patterns, which is insensitive to the overall rule size.

## III. CIRCUIT DESIGN AND IMPLEMENTATION

Specialized CAM circuits are designed to enable an arbitrary range of the array to be searched for the proposed three-phase architecture and further to enhance the overall memory and energy efficiency. Fig. 3 presents the diagram of a PE in Phase-1, which is composed of dual-port CAM with range activation as the match table and dual-port SRAM as the transition table. Phase-2 operates in a similar manner with a PE in Phase-1, except that it demands wildcard matching.

An 8-T NOR-type CAM, inspired by the 4+2T single-port BCAM cell in [11], is designed to act as dual-port BCAM for exact matching in Phase-1 and single-port TCAM for wildcard matching in Phase-2. The extended functionality is made possible by a fixed-1s data encoding scheme. Area and power savings are thus achieved over traditional single-port 10-T BCAM and 16-T TCAM cell while maintaining the same functionality. As shown in Fig. 7, the 8-T CAM cell with two standard access transistors for writing and two extra transistors with source-controlled search lines (SLs) for matching. Read disturbance caused by BL discharge during search is avoided, which troubles 6-T CAM cell implementation [16]. The 8-T cells ensure robust writing in any technology. A 4+2T cell can be used if the body bias effect on transistors' threshold voltage is strong as in [11]. The access transistors can be eliminated in such technologies.

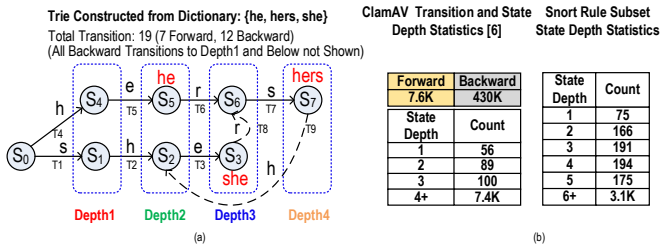

Fig. 4. (a) An example of Aho-Corasick algorithm Trie, (b) ClamAV and Snort selected sub-ruleset transition and state depth statistics.

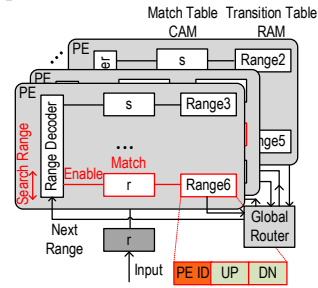

Fig. 6. Proposed pipelined range-matching architecture.

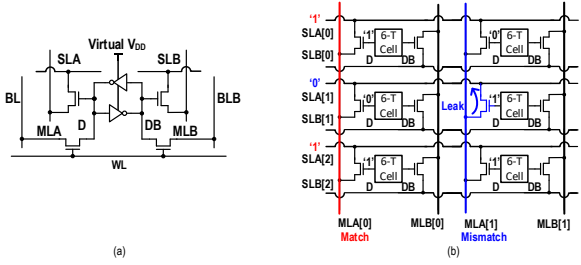

Fig. 7. (a) Proposed 8-T CAM Cell, (b) example search illustration, search "101" at port A, column 0 stores "101", column 1 stores "011".

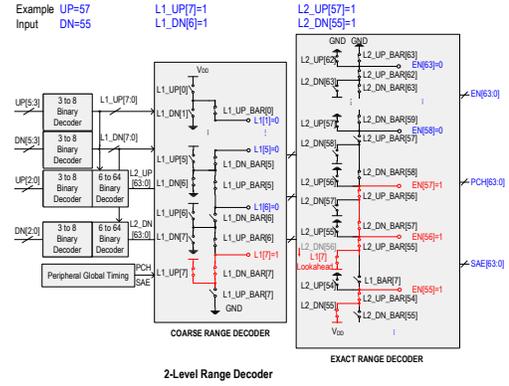

Fig. 9. 2-level range decoder and hierarchical clock gating.

Normal 8-bit characters in ASCII code from the pattern are encoded to 11b fixed-1s code, each with five '1's and six '0's at different locations (Fig. 8a). The total number of '1's and '0's of each code is the same: only positions are different. Based on the pigeonhole principle, mismatch between the encoded search pattern and stored string will always lead to at least one low SL ('0') and high stored Data ('1') pair (Fig. 8b), forming a leakage path to the precharged ML (Fig. 7). In this search scheme, only one SL and search transistor are necessary. An 8-T cell with two search transistors can perform a dual-port search with Port B searching inverted input patterns. As a result, throughput, memory efficiency, and energy efficiency are almost doubled at the cost of using 11 bits to represent an 8-bit character.

Fine-grained row enabling selects a range of rows to search and keeps the rest power gated to the retention voltage to reduce static power and limit short-circuit power caused by shorted SLs in certain scenarios (Fig. 7b). This shorting path between SLs is a concern for power but does not significantly reduce search reliability because the search transistors are NMOS. Extensive Monte-Carlo simulations confirmed a >450mV sensing margin.

The core circuit for row enabling is a hierarchical range decoder (Fig. 9) based on switches, which converts the binary range (UP/DN in Fig. 6) into enable signals for all selected match lines within bounded decoding time. The switch network is divided into segments by the decoded one-hot UP and DN switches, and the boundary switches (L1_UP in level-1 and L2_DN in level-2) activate the enable and timing signals in the selected range. For large ranges, level-1 decoder uses MSBs for a coarse range and accelerates the activation of level-2 by looking ahead for every eight rows, avoiding large RC delays caused by serially connected transistors and large parasitic capacitance when a large range is selected. The enabled range is exclusive of DN for level-1 and inclusive for level-2, leading to slightly different structures.

The proposed CAM cell and encoding scheme also support wildcard matching in Phase-2. Storing eleven '0's in the CAM will match all inputs and thus represents a "don't care" character as in TCAM (Fig. 8c). This approach can only be done on Port A and thus the other search port can be removed, resulting in a 7-T cell for single-port wildcard matching that replaces traditional 16-T TCAM cells. The associate SRAM in Phase-2 stores the matched pattern ID as well as the length of the matched patterns to facilitate clock gating after matching.

The Phase-2 search has a single port and takes two cycles, which is slower than the single-cycle and dual-port search in Phase-1. Data congestion can thus occur when Phase-1 reports intensive matches. Congestion includes two situations: two prefix-matches from Phase-1 in two consecutive cycles, and two prefix-matches from Phase-1 from two ports simultaneously. An arbiter is developed to handle various situations of congestion and higher priority is given to port A.

IV. MEASUREMENTS

A prototype chip is fabricated in 65-nm CMOS LP process. The chip micrograph is shown in Fig. 10a. In this prototype chip, a separate SRAM for Stage1 and eight PEs are implemented for Phase-1. Each Phase-1 PE contains 64 rows of 8-T CAM and SRAM. Phase-2 includes four 64x220 7-T CAM banks and associate SRAM banks. A 240-rule Snort subset is used for testing. When a rule pattern is found, the engine will report a "hit" and push the pattern ID to a buffer for Phase-3 processing in host processor. Input packets are randomly generated with controlled pattern hit rate for energy measurements.

For fair comparisons with other designs, energy efficiency is defined as the energy per search divided by the total number of characters in the ruleset ("Char"). The energy efficiency varies from 1.54 to 2.1 fJ/Char/search at 1.2 V and 144 MHz depending on pattern hit rate (Fig. 11a). The more the malicious patterns in packets, the more frequently Phase-2 will be activated. Phase-2 banks consume higher overall energy for a single search due to their larger capacity. Fig. 11b confirms that the energy consumption also depends on Phase-1 configurations and the plausibility of reconfigurability. Because of the higher Phase-2 energy, it is beneficial to avoid Phase-2 searches by including more stages in Phase-1. However, when the percentage of malicious packet rises above a certain threshold, it becomes more efficient to reduce the prefix depth of Phase-1 and pass more packets

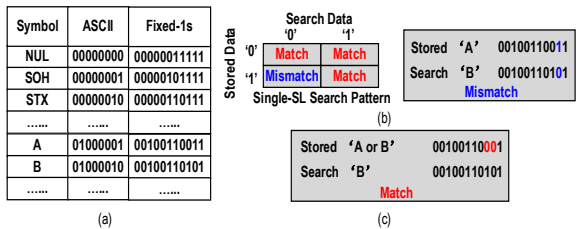

Fig. 8. (a) Encoding from ASCII code to fixed-1s code, (b) single side search example, (c) masking example under single side search.

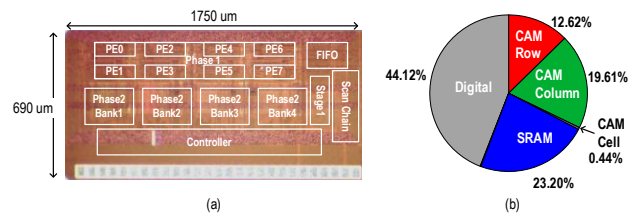

Fig. 10. (a) Chip micrograph, (b) Power breakdown at full workload.

TABLE I
COMPARISON TABLE WITH PRIOR ARTS

| | This Work | HPCA 2020 [6] | MICRO 2017 [7] | JSSC 2009 [9] | ISCA 2005 [3] | ICNP 2004 [8] | ICNP 2006 [10] | ASSCC 2019 [17] | VLSI 2018 [14] | JSSC 2018 [15] | JSSC 2004 [13] | JSSC 2005 [12] |
|---|---|---|---|---|---|---|---|---|---|---|---|---|
| Application | Payload Match | Payload Match | Payload Match | Payload Match | Payload Match | Payload Match | Payload Match | General Purpose CAM | General Purpose CAM | Header Match | Header Match | Header Match |
| Architecture | Pipeline Selective Enable | AC SRAM | AC SRAM | Wu-Manber BiTCAM | Bit-Split FSM SRAM | Brute-Force TCAM | AC, Transition Compression TCAM | 6-T CAM ML Clamping, Footer | Half and Half Compare TCAM | Don't Care Reduction CAM | Pipelined Hierarchical CAM | Hierarchical CAM |
| Technology | 65nm | 14nm | 28nm | 130nm | 130nm[a] | 65nm[b] | 65nm[b] | 28nm | 14nm | 65nm | 180nm | 100nm |
| Frequency (MHz) | 144 | 5000 | 1200 | 380 | 720 | 250 | 250 | 369 | 1470 | 330 | 143 | 300 |
| Throughput (MBps) | 288 | 80000 | 9400 | 380 | 720 | 250 | 250 | 369 | 1470 | 330 | 143 | 300 |
| Latency[f] (ns) | 41.7 | 4 | 16.7 | 5.3 | 27.8 | 4 | 80 | 2.708 | 0.68 | 3 | 7 | 3.3 |
| Power (mW) | 2.27 | 22000 | 1080 | 131.22 | 4720 | - | - | 0.066 (10 MHz) | 0.032 (10 MHz) | 46.7 | 60.8 | 31.1 |
| Memory Efficiency[c] (Byte/Char) | ~0.9 CAM ~0.6 SRAM | 12.8 SRAM | 17.2 SRAM | 0.66 CAM 0.62 SRAM | 18 SRAM | 2.5 CAM 20.6 SRAM | 4.75 CAM 2 SRAM | 1 | 1 CAM | 0.34 CAM | 1 CAM | 1 CAM |
| CAM Energy Efficiency[d] (fJ/Char/search) | 0.61 | - | - | 27 | - | 15.4 | 32 | 12.96 (10 MHz) | 3.04 (10 MHz) | 3.28 | 23.12 | 5.6 |
| DPI Energy Efficiency[e] (fJ/Char/search) | 1.54 | 22 | 26.8 | 43.2 | 293.4 | 40.4 | 44.5 | - | - | - | - | - |
| Cell Area (um²) | 1.33 | - | - | 10.32 | - | - | - | 0.362 | 1.47 | ~5.37 | - | 22.4 |

a. Modeled by [9]
b. Modeled using the technology by this work
c. Defined as total CAM byte on chip/total pattern bytes
d. Defined as energy of CAM/total pattern bytes per search
e. Defined as energy of system/total pattern bytes per search
f. Defined as delay time from input to output of an exact match

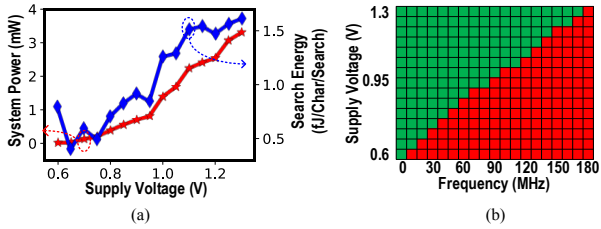

Fig. 11. (a) System power and search energy versus supply voltage (b) Shmoo plot of the system.

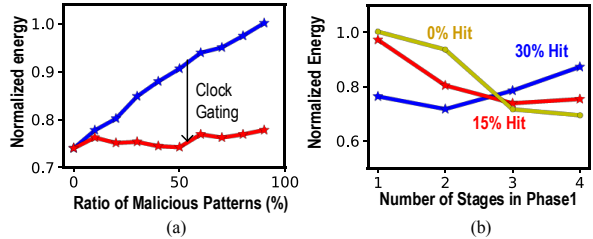

Fig. 12. (a) Normalized search energy versus ratio of malicious patterns (hit rate) with and without clock gating, (b) Normalized search energy versus number of stages in Phase-1 under different attack rate.

to Phase-2. Moreover, Fig. 11a shows that clock gating after hits saves up to 25% power and renders the overall efficiency less dependent on the hit rate. Fig. 10b presents the power breakdown in one scenario, when both ports are activated, Phase-1 is configured with four stages, and an input stream with 90% malicious packets. Fig. 12a plots the chip's voltage dependence of power consumption. The system shmoo plot is shown in Fig. 12b. Comparisons with related prior arts, including accelerators for DPI, header matching, and generic TCAMs, are listed in Table. I.

## V. CONCLUSION

This paper presents an energy- and memory- efficient CAM-based payload matching engine for NIDS in IoT [18]. Leveraging holistically designed architecture and circuits, a 65-nm prototype achieves best-in-class 1.54-fJ/search/pattern byte and 0.9-byte/pattern byte.